\documentclass{elsart}

\usepackage{epsfig}

\usepackage{amssymb}

\begin{document}

\begin{frontmatter}

\title{Modeling the Searching Behavior of Social Monkeys}

\author[IFUNAM]{D. Boyer\corauthref{cor}},
\corauth[cor]{Corresponding author}
\ead{boyer@fisica.unam.mx}
\author[IFUNAM]{O. Miramontes},
\author[MEDA]{G. Ramos-Fern\'andez},
\author[IFUNAM]{J.L. Mateos},
\author[IFUNAM]{G. Cocho}

\address[IFUNAM]{Departamento de Sistemas
Complejos, Instituto de F\'\i sica, Universidad
Nacional Aut\'onoma de M\'exico, Apartado Postal 20-364, 
01000 M\'exico D.F., M\'exico}

\address[MEDA]{Pronatura Pen\'\i nsula de Yucat\'an, Calle 17 $\#$ 188A x 10,
Col, Garc\'\i a Giner\'es, M\'erida, 97070 Yucat\'an, M\'exico}

\begin{abstract}
We discuss various features of the trajectories of spider monkeys looking
for food in a tropical forest, as observed recently in an extensive 
{\it in situ} study.
Some of the features observed can be interpreted as the result of 
social interactions. In addition, a simple model of deterministic walk in a 
random environment reproduces the observed
angular correlations between successive steps,
and in some cases, the emergence of L\'evy distributions for the length 
of the steps.
\end{abstract}


\end{frontmatter}

\section{Introduction}\label{intro}

It is often hypothesized that animals live in groups because they benefit 
from predator avoidance and collective food searching \cite{krebs93}. However, 
group living also imposes behavioral restrictions on individuals and may,
on the other hand, 
facilitate the emergence of entirely new behavioral traits. 
Foraging (or food searching, preparation and consumption) is one of the 
best-known examples of behavior modulated by group. 
Recently, several studies on various species have shown that lone foraging 
animals follow trajectories characterized by L\'evy probability 
density distributions for the step lengths 
\cite{levandowsky88,viswanathan96,viswanathan99}. 
In physics, L\'evy flights and walks are a manifestation
of anomalous diffusion (see \cite{shlesinger93,klafter96} for reviews).
It has been suggested that L\'evy distributions in animals lead to an 
optimal search \cite{viswanathan99,viswanathan00,viswanathan02}.

In a previous study \cite{ramos03},
it was found that grouping has indeed an impact on the searching patterns 
of the social spider monkeys ({\it Ateles geoffroyi}). The lengths of 
traveled distances between two stops
follow a L\'evy-like distribution with a scaling exponent that is different 
for individuals in groups and individuals when occasionally alone. 
This difference in the exponent values evidences subtle differences in 
movement patterns over long distances. The foregoing result
suggests that collective 
searching may be indeed a more efficient strategy, since the exponent value 
is close to what has been argued elsewhere to be the exponent 
for an optimal L\'evy searching process \cite{viswanathan99}. In the following,
we present probabilistic arguments to quantify group effects during foraging
(Section \ref{group}). In Section \ref{greedy},
a simple deterministic walk model in a random environment
reproduces some of the foraging patterns observed on spider monkeys.

\section{Searching and effective group size}\label{group}

Spider monkeys forage in subgroups that change in size and composition several
times during the day \cite{klein77}. These subgroups remain coherent (i.e. with
individuals separated by no more than 30 meters from each other) for 
several hours,
before they split or are joined by other 
group members \cite{krebs93,kummer68}. 
These subgroups may contain from one to eight adults
and their young, the majority containing two adults \cite{ramos03b}.
It has been suggested that subgroup size varies in response to the size of the
food patches where spider monkeys feed, 
and that they may share information on the presence of newly found food
patches \cite{chapman90}.
In a recent study \cite{ramos03}, different food searching patterns were 
observed 
between subgroups containing one adult and those containing more than one.
In particular, the difference laid in the distribution ($P(l)$) of step 
lengths, defined as the probability that a monkey traveled a distance 
$l$ in a fixed time interval (5 minutes). 
The measured
distributions for lone adults and larger subgroups, $P_{s}(l)$ and $P_g(l)$ 
respectively, can both be fitted by L\'evy laws 
($P_i(l)\sim l^{-\alpha_i}$, $i=s,g$), with different exponents. 
The data are consistent with $\alpha_s\simeq 1.5$ and $\alpha_g\simeq 2.1$,
for lone and grouped individuals, respectively \cite{ramos03}.

We develop below a possible explanation for these results.
Each individual in a subgroup has a particular knowledge of  
the areas of the forest where food is likely to be found. 
This knowledge is not ideal and
may vary from one monkey to the other, since each one
has its own searching history (accumulated over days or weeks
of foraging alone or with different combinations of group members). 
In a given subgroup, this knowledge is likely to be shared. Moreover,
it is often observed that, during motion, the subgroup is decomposed 
into individuals that are actively searching, while the others follow rather 
passively. Sometimes, active individuals 
follow straight lines between food patches, as if
they knew where they were going. 
Yet, there are no invariant leaders; the identity 
of active monkeys continuously change over time, as sketched in 
Figure 1. 
On the other hand, lone individuals cannot profit from the cooperation of 
other monkeys, thereby reducing their searching efficiency, resulting in 
longer trajectories ({\it i.e.} $\alpha_s<\alpha_g$, as observed). 

\begin{figure}\label{fig:group}
\begin{center}
\epsfig{figure=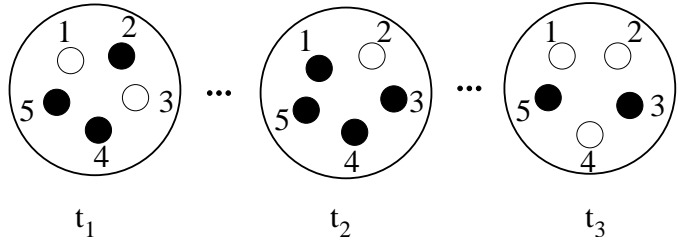,width=3.5in}
\end{center}
\caption{A subgroup of 5 monkeys composed of active ($\circ$) 
and passive ($\bullet$) individuals during the search process.  
The identities and number ($n^*$) of active individuals may change during 
the course of time.}
\end{figure}

We propose a simplified probabilistic argument to quantify these ideas. 
Suppose that the average
intrinsic (and imperfect) knowledge that each monkey has of its environment 
can be translated into the \lq\lq lone" step length distribution 
$P_s(l)\sim l^{-\alpha_s}$. 
Assume that in a subgroup of $n$ monkeys, there are in average 
$n^*\le n$ actively searching individuals. At time $t=0$, the subgroup starts 
to move coherently in a given direction. Each active individual $i$
would {\it a priori} travel a distance $l_i$, chosen independently from
the subgroup according to the distribution $P_s$. 
The subgroup stops when one of the $n*$ active individuals first decides so 
({\it e.g.} it finds food that no one else has seen and calls all others
in the subgroup). The length $l_g$ 
traveled by the subgroup is then given by
$l_g={\rm Min}(l_1,..,l_{n^*})$.
The trial steps being assumed independent, the distribution function
$P_g$ of $l_g$ obeys: 
\begin{equation}
\int_{l_g}^{\infty}P_g(l)dl=\left(\int_{l_g}^{\infty}P_s(l)dl\right)^{n^*}
\end{equation}
If $P_s$ is a L\'evy distribution, so is $P_g$ and its exponent is 
$\alpha_g=n^*(\alpha_s-1)+1$. Given the exponents measured experimentally
($\alpha_g\simeq 2.1$, $\alpha_s=1.5$ \cite{ramos03}), this gives a value 
$n^*=2.2$.
(Note that $n^*$ is an average number, it does not have to be an integer.)
This rather small value is very close to the modal subgroup size of 
two adults per subgroup.
The higher $n^*$, the slowest
the motion of the subgroup. Too much active individuals would result in
a Gaussian motion (a decay faster than $l^{-3}$), and therefore a
limited exploration of the territory.
We conjecture that a good balance between cooperation during foraging
($n^*$ large enough), a good exploration of the territory 
(L\'evy-like, {\it i.e.} $n^*$ not too large),
and a low level of competition for food ($n$ not too large),
is maintained in subgroup of small sizes.

\section{A \lq\lq greedy" deterministic walk model}\label{greedy}

In recent approaches of food search problems, animal movements
are modeled by stochastic random walks characterized by L\'evy 
distributions given {\it a priori} \cite{viswanathan99,viswanathan00}.
Here, we rather present an approach
where animal trajectories are strongly coupled to (or induced by)
a complex spatial distribution of resources. The aim is to understand 
the effects of a random spatial
distribution of fixed resources on monkey trajectories.
In the experimental study \cite{ramos03} , monkeys used regular routes to 
travel between feeding sites, within a limited area of 
about $2$ km$^2$ \cite{ramos02}. 
Many animals (bees \cite{dyer94}, rodents \cite{collett86},
primates \cite{garber89,janson98}) seem to rely on cognitive maps 
in order to navigate their environment. 
These maps may contain information on the location 
of different targets and the geometric relationships 
between them \cite{kamil97}. Some species of monkeys can also detect 
the closest target, as observed experimentally in \cite{janson97}.
Various animals actually keep in 
memory the sites that they have already visited in a recent past
\cite{garber89,kamil97,dyer94}. 
Models focusing on geometrical detection processes have been 
proposed (see \cite{janson98} for an overview), but statistical
analyzes of are still scarce.

Consider a two-dimensional space composed of point-like targets
randomly distributed in space, representing the trees where
monkeys can find fruits. 
\begin{figure}\label{fig:angle}
\vspace{-3.5cm}
\begin{center}
\hspace{-0.5cm}\epsfig{figure=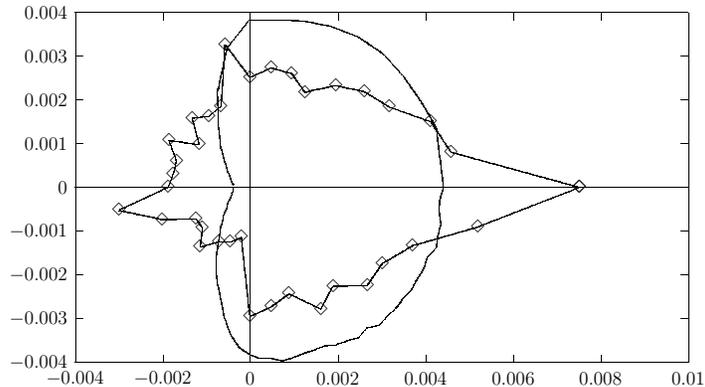,width=5.in}
\vspace{-9.cm}
\caption{Normalized polar plot of the angular distribution $p(\theta)$
of the turning angle between consecutive steps ($\theta$ in degrees),
as measured in the experiments (dots), and obtained in the greedy model 
(solid line). Simulations are made with $N=10^6$ targets in a square
domain, averages are performed over 50 independent runs.}
\end{center}
\end{figure}
The rules of motion are the following (see ref. \cite{lima01} for
a related model): $(a)$
a monkey located at the target number $i$ will next move in straight line
to a target $j$ such that $l_{ij}$ is minimal among all targets, 
where $l_{ij}$ is the distance separating $i$ and $j$;
$(b)$ The monkey does not jump to an already visited target. 
This parameter-free \lq\lq greedy" model generates random walks
due to the random positions of the targets. It exhibits
angular correlations between successive moves, a 
feature observed in real monkey trajectories. Figure 2 
shows the
probability distribution $p(\theta)$ that a walker turns an angle $\theta$
with respect to its previous step, obtained from numerical solutions. 
With the condition $(b)$, it is
more likely that the walker encounters unvisited sites in front of him
than behind him: $p(\theta)$ peaks around zero. A qualitative agreement
is obtained between the observed angle distribution and
that of the model. However,
the model overestimates the moves toward the sides, and under-estimates 
U turns ($\theta\sim180^{\circ}$). 

\begin{figure}\label{fig:pdf}

\vspace{-3.cm}
\begin{center}
\hspace{-0.5cm}\epsfig{figure=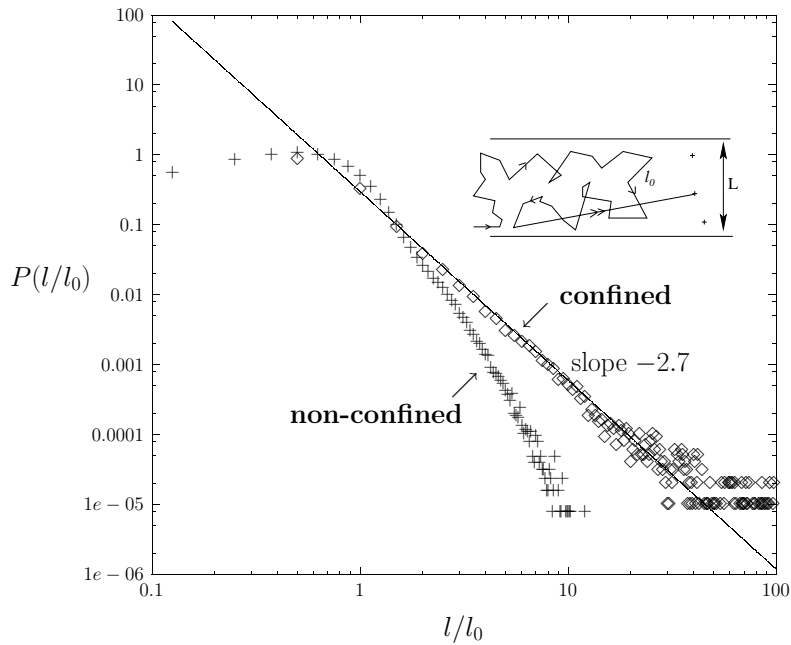,width=5.in}
\vspace{-6.5cm}
\caption{Normalized step length distribution $P(l)$ of the greedy model,
obtained with $N=10^6$ targets in a domain of area $1$, averaged over
10 runs (when each run stops, the number of visited sites is still $\ll N$);
$l_0=N^{-1/2}$ is the average distance between 2 nearest targets. 
$(+)$: square (unbounded-like) domain; $(\diamond)$: narrow domain of width 
$L=10l_0$ (see inset). The solid line is a guide to the eye.}
\end{center}
\end{figure}

The simplest version of the model can not account for the L\'evy walks
observed in the field. We hence propose the following modification.
The home range of spider monkeys is actually not an infinite plane, 
but covers a fairly narrow and stretched area \cite{ramos02}.
With the rules of the greedy model, we now consider targets 
distributed within a narrow strip, of width $L$ and infinite in length.
Figure 3 displays the distribution of the length of the steps 
$P(l)$ obtained numerically for such a configuration and 
$L=10 l_0$ (with $l_0$ the typical
distance between two nearest targets). The (exponential-like) distribution 
for the unbounded territory ($L\rightarrow\infty$) is also shown
for comparison. The confined trajectories have a strikingly broad
step distribution, consistent with a L\'evy law of
exponent close to $2.7$. This can be explained qualitatively as follows:
In the simulations, the
walker follows a nearly 1D path ({\it e.g.} moving toward the right in
average), but sometimes goes backward to eat some unvisited targets left
behind. After some time, these backward paths usually end up in a region
with no more fresh targets to eat, and these may be located quite far 
away from the most advanced point of the trajectory. To reach the
following nearest site, the walker then needs 
to jump back at the front with a long step (see inset in Fig.\ref{fig:pdf}). 
Surprisingly, it seems that such  
broad distributions are observed most easily for $L/l_0\sim10$. 
Short range distributions are recovered as $L/l_0\rightarrow\infty$ or 
$L/l_0\rightarrow 1$.

\section{Conclusion}\label{concl}

We have discussed several aspects of foraging patterns of social monkey
in a well defined territory. 
Foraging in small groups can modify searching patterns provided
that monkeys cooperate by sharing information on the location of food
patches.
In addition, we have discussed a simple model
based on deterministic rules of motion, emphasizing a strong coupling between 
the animals and the statistical distribution of resources. 
The results reproduce qualitatively some of the experimental data
(angular correlations, and in some case,
trajectories with L\'evy distributions).
Future improvements should incorporate 
the temporal variation in the location and
abundance of food patches, as well as the consequences of 
foraging patterns for the dispersion of seeds by fruit-eating monkeys.

{\bf Acknowledgements}

We are indebted to F. Leyvraz for illuminating discussions.
This work was supported by CONACYT grants G32723-E and C01-40867,
DGAPA grant IN-111000, as well as a visiting scholarship
(C\'atedra Tom\'as Brody) from the Complex Systems Department
at the Physics Institute, National Autonomous University of Mexico.


\end{document}